\documentclass[runningheads]{llncs}
\usepackage{marvosym}
\usepackage{orcidlink}
\usepackage[T1]{fontenc}
\usepackage{graphicx}
\usepackage{hyperref}
\usepackage{color}

\begin{document}
\title{From Intention to Text: AI-Supported Goal Setting in Academic Writing}
%
%
\author{Yueling Fan\inst{1}\orcidlink{0009-0004-0737-9196} \and
Richard Lee Davis\inst{2,3}\orcidlink{0000-0002-6175-9200} \and
Olga Viberg\inst{1,3}\textsuperscript{(\Letter)}\orcidlink{0000-0002-8543-3774}}
\authorrunning{Y. Fan et al.}
%
\institute{Department of Media Technology and Interaction Design, KTH Royal Institute of Technology, Stockholm, Sweden  
\\ \email{yueling@kth.se} \and
Department of Digital Learning, KTH Royal Institute of Technology, Stockholm, Sweden \and
Digital Futures, Stockholm, Sweden \\
\email{\{rldavis,oviberg\}@kth.se}}
\maketitle              
\begin{abstract}
This study presents WriteFlow, an AI voice-based writing assistant designed to support reflective academic writing through goal-oriented interaction. Academic writing involves iterative reflection and evolving goal regulation, yet prior research and a formative study with 17 participants show that writers often struggle to articulate and manage changing goals. While commonly used AI writing tools emphasize efficiency, they offer limited support for metacognition and writer agency. WriteFlow frames AI interaction as a dialogic space for ongoing goal articulation, monitoring, and negotiation grounded in writers’ intentions. Findings from a Wizard-of-Oz study with 12 expert users show that WriteFlow scaffolds metacognitive regulation and \textit{reflection-in-action} by supporting iterative goal refinement, maintaining goal–text alignment during drafting, and prompting evaluation of goal fulfillment. We discuss design implications for AI writing systems that prioritize reflective dialogue, flexible goal structures, and multi-perspective feedback to support intentional and agentic writing.

\end{abstract}
\section{Introduction}
Academic writing is a cornerstone of higher education, serving not only as a medium for assessment but as a powerful engine for learning, knowledge construction, and intellectual development. Rather than  reproducing information, students engage in writing as a process of knowledge transformation, negotiating the dynamic interplay between rhetorical challenges (how to communicate ideas) and content-related challenges (what to communicate) \cite{bereiter1987attainable}. Through this process, tacit, experiential, and fragmented knowledge can be externalized, refined, and made transferable across contexts \cite{tynjala2001writing}. Academic writing further fosters strategic and self-regulated cognitive skills \cite{hayes1986writing}, including goal setting, planning, monitoring, and revising---capabilities that are critical for scholarly inquiry, professional practice, and lifelong learning. However, despite its importance many students struggle to meet the demands of academic writing \cite{pineteh_2013_the,sabti_2019_the}, highlighting a persistent gap between the recognized importance of writing and students’ ability to effectively engage in it.

With the rise of large language models (LLMs), the ways students engage in academic writing practices have started to change. For example, systems such as ChatGPT show strong reasoning and open-ended text generation capabilities \cite{xu2025large,becker2024text}, and have become increasingly embedded in students’ learning \cite{freeman2025student}. Yet, growing evidence suggests that reliance on such tools may undermine learning by encouraging cognitive offloading and reducing metacognitive engagement \cite{fan2025beware}. In academic writing, these risks are particularly acute, since its value lies not only in text production but in sustained reflection, reasoning, and iterative goal revision \cite{flower1981cognitive,hayes1986writing}.

Academic writing is a recursive and cognitively demanding process requiring writers to manage evolving goals, integrate cross-sectional ideas, and continually evaluate and revise arguments \cite{flower1981cognitive,hayes1986writing}. However, most commercial LLM interfaces are optimized for linear, turn-based dialogue, which poorly aligns with non-linear writing processes. Revisiting earlier reasoning or managing multiple concurrent goals is cumbersome, often resulting in surface-level interactions that limit metacognitive regulation, which is a strong predictor of academic success \cite{viberg2020self}.

These challenges can be seen through the lens of self-regulated learning (SRL), which emphasizes learners' active metacognitive regulation across forethought, performance, and reflection phases \cite{schunk2011handbook}, with goal setting as a central mechanism \cite{zimmerman2002becoming}. In academic writing, goals are continuously revised as ideas evolve, making dynamic goal regulation essential for preserving the epistemic value of writing with AI. 

Recent AI-driven writing tools have begun incorporating SRL lens to support self-reflection and critical evaluation \cite{li2025turning,wong2024supporting}, yet do not explicitly support \textit{iterative goal adjustment}.
This study aims to fill this gap by examining how AI-supported systems can scaffold reflection and goal setting in academic writing. We present \textit{\textbf{WriteFlow}}, a Google Docs–based writing assistant co-designed to support reflection-in-action \cite{schon2017reflective} through AI-mediated dialogue, structured goal generation, and goal-alignment tracking. Building on prior work showing conversational interaction as a powerful design resource for fostering reflection \cite{bentvelzen2022revisiting}, WriteFlow reconceptualizes chat-based interaction as a space for goal negotiation and metacognitive regulation.

This study contributes to research on AI-supported academic writing by (1) providing a formative account of students' writing goal–related stress and current patterns of LLM use; (2) introducing \textit{WriteFlow}, a voice-based AI writing assistant that scaffolds metacognition and self-regulation through iterative construction and monitoring of writing goals; (3) providing empirical evidence of how WriteFlow supports tracking alignment between stated goals and emerging text; and (4) deriving design implications for human–AI writing systems that support evolving goal setting and metacognitive scaffolding.

\section{Background}

\subsection{Goal Setting for Self-Regulated Academic Writing}

Academic writing is guided by hierarchically structured goal structures that include abstract intentions (e.g., audience awareness) and concrete subgoals (e.g., revising a paragraph) \cite{flower1981cognitive,hayes1986writing}. Expert writers generate richer goal structures and flexibly revise them as task constraints evolve \cite{hayes1986writing}. Research shows planning and quality-oriented goals improve text quality and promote higher-level revision behavior \cite{beauvais2011some}. Specific, proximal, and appropriately challenging goals enhance metacognition, motivation, and overall writing performance \cite{schunk1990goal,reid2013strategy,chung2021impact}. Process-oriented goals are particularly effective, compared to product-focused goals, process goals combined with feedback support learning of writing strategies, strengthen self-efficacy, and promote transfer \cite{schunk1990goal}. Although automated writing evaluation systems provide personalized feedback \cite{jansen2024individualizing}, they offer limited support for reflection on writers' underlying intentions or monitoring evolving goal structures during the writing process.

\subsection{AI Writing Tools and Metacognitive Support}

Recent AI writing tools have increasingly incorporated support for metacognitive processes during academic writing. Systems such as VISAR \cite{zhang2023visar} enable writers to construct hierarchical goal structures during the planning phase, helping them articulate and organize content goals before drafting begins, though research indicates such interfaces can increase cognitive load \cite{radwan2024sard}. Other tools focus on supporting metacognitive reflection during revision and feedback stages. Friction \cite{zhang2025friction} helps writers formulate actionable revision goals when editing existing drafts, while ALure \cite{neshaei2025leveraging} scaffolds self-regulated learning through structured prompts that encourage writers to reflect on their strategies and progress. Reverse outlining approaches \cite{dang2022beyond} enable retrospective assessment of whether written text aligns with intended structure, promoting reflective revision. Research on feedback timing suggests that continuous, in-action feedback better supports learning than post-hoc evaluation alone \cite{karolus2023your,yen2024give}, though such approaches also risk fostering dependency if not carefully designed.

Despite these advances in metacognitive support in planning and revision stages, existing tools do not help writers track the evolving relationship between their stated goals and emerging text during the drafting process itself. Writers lack explicit mechanisms to notice when drift occurs, i.e., when the text being produced no longer serves the goals originally intended. This gap is particularly acute in AI-assisted writing contexts, where generated content may subtly pull writers away from their intentions without them recognizing the misalignment until substantial revision is required. Writers need support not only for setting goals (planning tools) and evaluating completed text (revision tools), but also for maintaining awareness of goal-text alignment throughout drafting, enabling them to decide whether to realign their text with original goals or intentionally revise those goals in light of emerging insights.

Addressing this gap, we present empirical findings on how an AI-based writing assistant can be designed to support metacognitive scaffolding through iterative goal setting and monitoring. This study consists of (1) a formative study informing the co-design of WriteFlow and (2) an expert user evaluation.

\section{Formative Study}
To understand how adult writers use AI when addressing academic writing challenges, we conducted a survey with 17 students (8 female, 8 male, 1 non-binary; aged 22–34). The open-ended online survey\footnote{Survey instrument, participant demographics, and detailed findings for the formative study: OSF repository, \url{https://osf.io/ba6d2/overview?view_only=e69b00a3acfd42529d3fb2c9c10b1ef7}} was administered between May 22 and 27, 2025 and included ratings of ten writing challenges grounded in Cognitive Process Theory of Writing which posits that writing is a non-linear, goal-directed, and recursive mental process rather than a strictly staged product \cite{hayes1986writing}, along with questions about coping strategies and AI use. Responses were analyzed using Reflexive Thematic Analysis \cite{clarke2014thematic}.
The study identified three key writing challenges and 20 coping strategies (detailed descriptive statistics and thematic analysis results are available in our OSF repository\footnote{Survey instrument, participant demographics, and detailed findings for the formative study: OSF repository, \url{https://osf.io/ba6d2/overview?view_only=e69b00a3acfd42529d3fb2c9c10b1ef7}}). The rating data showed that the most stressful challenges were \textit{evolving writing goals} and \textit{setting writing goals}, reflecting difficulties in revising plans as new ideas and sources emerge. Participants reported using outlining, documentation, and AI tools to track evolving ideas, test structural changes, and maintain alignment with central arguments. These strategies supported reflection and reduced uncertainty during revision. In their open-ended responses, the study participants also highlighted a third key challenge: \textit{preserving authorship}. Some participants (n = 6) reported primarily using ChatGPT in a \textit{human-in-the-loop} manner, leveraging it for ideation, tone refinement, and feedback interpretation. However, other participants (n = 9) expressed concerns about overreliance, authenticity, and ownership. Based on these findings, we derived five design requirements (R) for AI-supported reflective academic writing: (\textbf{R1}) facilitate goal articulation; (\textbf{R2}) support iterative goal refinement; (\textbf{R3}) enable organization and revisiting of ideas relative to goals; (\textbf{R4}) preserve writer voice and meaning; and (\textbf{R5}) ensure AI feedback is transparent, revisable, and aligned with user intent. Together, these requirements underscore the centrality of human judgment and agency in AI-assisted academic writing.

\begin{figure}
    \includegraphics[width=1\textwidth]{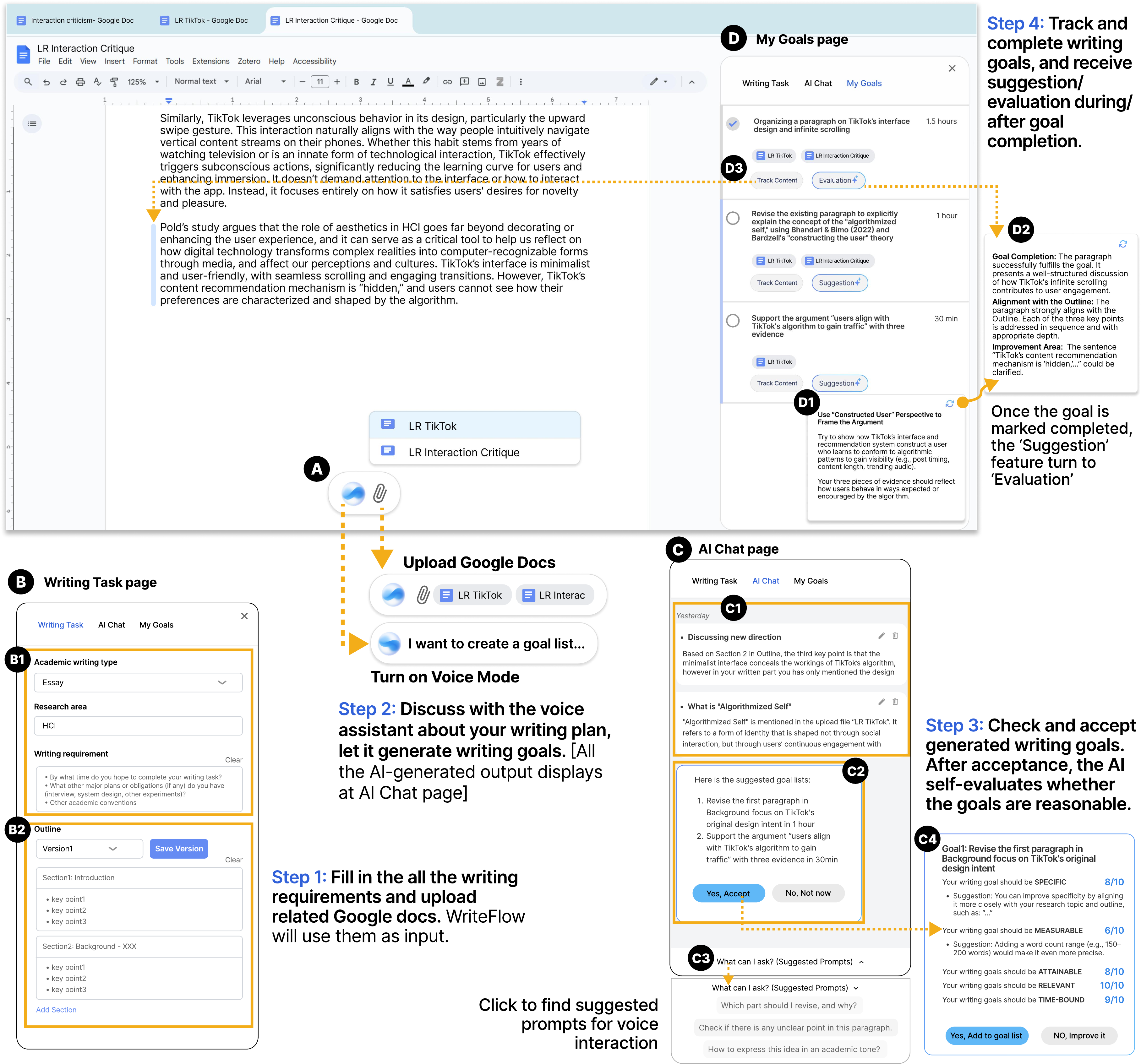}
    \caption{The overview of WriteFlow, a Google Docs add-on for goal-oriented academic writing. WriteFlow interface consists of a voice agent (A) and a sidebar panel with three pages: Writing Task (B), AI Chat (C), and My Goals (D). Users can upload Google Docs and communicate with the voice agent at any stage of writing to discuss their writing directions. The agent then generates writing goals to help them plan, track, and monitor their writing process.}
    \label{fig:The system overview of WriteFlow}
\end{figure}

\section{System Overview}\label{sec3} Figure 1 presents WriteFlow's workflow and interface \footnote{WriteFlow ProtoPie prototype: \url{https://cloud.protopie.io/p/3176a8c0ab9ad1f9e99b0910}}. Users provide writing requirements and upload drafts, which the system uses to contextualize task understanding. Through Voice Mode, users discuss their writing plans with an AI-mediated conversational agent, which generates writing goals aligned with their intentions. Goals are stored on the\textbf{ My Goals} page, where users can track progress, receive targeted suggestions, and review post-completion evaluations. WriteFlow also provides an \textit{Outline} view that supports creating, revising, and comparing multiple outline versions across drafting stages.

\textbf{Goal Setting and Monitoring.} WriteFlow supports planning and self-regulated writing by scaffolding goal articulation, refinement, and progress monitoring. During voice-based interaction, the system helps users externalize ideas and translates them into adaptive writing goals (\textbf{R1}, \textbf{R3}). Self-evaluation cards enable users to iteratively revise goals (\textbf{R2}), while progress tracking (\textbf{R5}) and suggestion cards support focused execution and sub-goal formation. 

\textbf{Goal-Text Alignment Evaluation.} Central to WriteFlow's design is the Goal Completion Evaluation feature, which directly addresses the challenge of tracking alignment between evolving goals and emerging text. After goal completion, the system evaluates alignment between goals, outlines, and written content to support reflective revision and preservation of authorial intent (\textbf{R4}). The Outline view further supports flexible goal evolution by allowing users to create and compare multiple outline versions across drafting stages, enabling tracking of how writing plans change over time.

\section{User Evaluation}
We used WriteFlow as a design probe and conducted an exploratory Wizard-of-Oz study to answer the following \textbf{research questions}: \textbf{RQ1.} In what ways does the use of WriteFlow support users' metacognitive goal-oriented processes during academic writing, and what design refinements are needed, based on their feedback? \textbf{RQ2.} How does using WriteFlow influence writers’ sense of agency and their reliance on AI during academic writing?





\subsection{Participants}

A total of 12 writers with expertise in human–computer interaction (HCI) participated in the study (9 female, 3 male; ages 23–29). All had experience in English academic writing, and 7 had prior academic publication experience in the HCI field. Participants were intentionally recruited from the HCI domain, as they represent a population for whom reflective academic writing is a routine yet demanding practice and who possess the analytical skills required to critically interrogate interactive system behavior. Their familiarity with academic writing conventions and the design and evaluation of interactive systems enabled an informed assessment of WriteFlow’s support for goal articulation, authorial voice, and user control. Also, participants exhibited diverse AI use practices: five reported very frequent use of AI for writing, four reported frequent but more limited use, and three reported occasional use for specific stages of the writing process. This variation allowed us to examine how prior experience with AI shaped engagement with and perceptions of the system during real-world writing tasks.


\subsection{Study Design}
Participants interacted with WriteFlow, a high-fidelity \href{https://www.protopie.io/}{ProtoPie} prototype supporting multimodal input (text, document upload, and simulated voice). Although presented as an autonomous AI assistant, all responses were controlled in real time by a human facilitator (Wizard-of-Oz). Participants were asked to imagine they were completing a realistic academic task: writing a TikTok interaction critique from a user experience perspective. All study materials, including the task instructions, prompts described below, are available in an OSF repository\footnote{\url{https://osf.io/xjw6f/overview?view_only=f7779496c1ba4bbab7c928767a5cd7f2}}.
WriteFlow
\textbf{Setup.} We used GPT-4o to generate all AI responses. All prompts were pre-defined and pilot tested to ensure consistency and task relevance. To help participants quickly engage with the writing task, we provided pre-prepared literature review notes on interaction criticism frameworks and TikTok case studies. Participants were encouraged to use these materials but could also incorporate their own ideas. We focused on how participants used WriteFlow during the writing process, particularly their strategy use and cognitive processes during the WriteFlow-assisted academic writing process, rather than evaluating final written outputs.

\subsection{Procedure}

\begin{figure}[t]
  \centering
  \includegraphics[width=\linewidth]{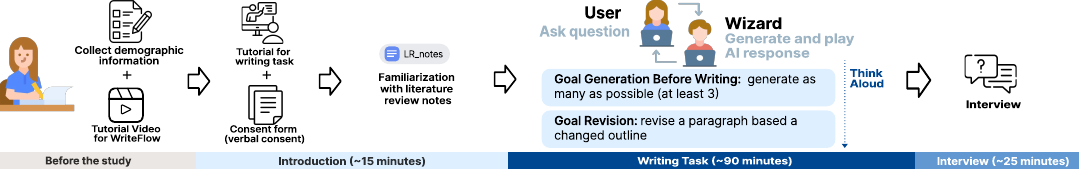}
  \caption{Procedure of the user study.}
\end{figure}

The study involved two writing tasks followed by post-study semi-structured interviews (Fig. 2). Each participant spent 120-150 minutes total and received a gift voucher as compensation for their time. All sessions were conducted remotely via Zoom and were both screen- and audio-recorded for accurate transcription. 
Before the study, participants provided demographic information and watched a brief video demonstrating WriteFlow's goal-setting workflow. The study duration (over 2 hours) enabled deeper interaction with the system. At the beginning of the study, participants completed a written consent form and were introduced to the writing assignment. They were given 10 minutes to familiarize themselves with the provided literature review notes. Participants completed two writing tasks while thinking aloud. In Task 1 (Goal Generation Before Writing), they discussed their initial ideas with the AI, which generated writing goals. They evaluated and chose to accept or reject the suggested goals (minimum three goals). In Task 2 (Goal Revision), participants identified misalignments between a pre-written paragraph and a revised outline. A semi-structured interview followed to understand participants' perceptions and experiences with WriteFlow.

\subsection{Data Analysis}
Recordings of users' (n = 12) interactions with WriteFlow (a total of 18 hours) and post-study semi-structured interviews (a total of 6 hours) were transcribed and analyzed using reflexive thematic analysis \cite{clarke2014thematic} by two researchers. We first familiarized ourselves with the transcripts by repeatedly reading and taking analytic notes on participants’ transcripts. Then we open-coded how participants experienced WriteFlow during their writing tasks and how it succeeded or failed to support writers' self-reflection. This coding was conducted in an inductive, data-driven manner, staying close to participants’ language and concrete experiences. After the first few rounds of coding and discussion, the authors observed that most of the issues participants focused on when evaluating whether WriteFlow was helpful closely resonated with the five design requirements (\textbf{R1–R5}). For example, in their think-aloud protocols, participants frequently commented on how specific features supported their efforts to articulate and refine goals, organize and reuse ideas in the notes, preserve their authorial voice, etc. Building on these observations, the first author then revisited the previously derived design requirements and we began to treat (\textbf{R1–R5}) as an analytic lens to attend more closely to how participants’ experiences aligned with, extended, or challenged each requirement. We also extended our coding to capture additional design implications and considerations that participants proposed. Subsequently, the researchers conceptualized a set of themes and met regularly to check for disconfirming cases, merge or split themes when needed. Ultimately, we present two main themes in the Results section.



\section{Results}

\subsection{Goal-Oriented Workflow as Metacognitive Scaffolding}
\textbf{Facilitated Goal Articulation (R1).} WriteFlow was experienced to support participants in clarifying "vague" or "underdeveloped" ideas (U1, U2, U4, U12) and generating "new perspectives" aligned with their writing intentions (U2, U4, U12). U12 for instance, appreciated voice input for enabling more fluid thinking, facilitating her initial idea formation: “I might say a lot of nonsense at first, but I become clearer and clearer in the process”. U11 further highlighted that "goal-setting helped define the scope of AI interaction within the broader writing task", making conversations more "focused" and "purposeful." Additionally, U3 and U8 suggested WriteFlow could be improved by asking follow-up questions or recommending relevant readings based on the ongoing conversation.

\textbf{Support Iterative Goal Refinement (R2).} Several writers reported that WriteFlow enabled them to flexibly and continuously adjust their writing plans throughout the writing process (U1, U2, U5, U6, U7, U9, U11). U2 explained, “I can easily modify my writing plan at any stage. I don’t need to be scared of losing the direction.” Similarly, U5 noted: “you can go through and filter or refine those goals yourself,” highlighting the WriteFlow’s support for iterative goal refinement.

U7 and U8 reported the \textit{Track Content} feature improved writing efficiency by automatically linking goals to relevant text segments. This function was particularly valuable during paper revision: “You’ll have many goals based on the reviewer’s feedback, each requiring changes in different parts” (U12). U2 similarly stressed the feature helped her quickly navigate longer papers with multiple goals. Additionally, three other participants expressed a desire for the ability to directly navigate to source paragraphs within goal-linked documents (U2, U8, U9).

Participants (n = 7) also emphasized the importance of visualizing the hierarchical relationships between writing goals, suggesting that such representations could reduce cognitive overload and support the development of a more coherent and comprehensive goal network. To further improve goal management, U8 proposed that completing a parent goal should automatically mark its sub-goals as 'completed', while others suggested categorizing goals by issue type (e.g., structure, citation) (U7) or by difficulty level (U11). U11 noted a preference for prioritizing more challenging goals first, indicating the value of flexible goal prioritization strategies.

\textbf{Enable organization and revisiting of ideas relative to goals (R3).} Goal completion evaluation appeared to enhance metacognitive awareness by prompting writers to assess whether their drafts fulfilled their intended goals. Participants reported that WriteFlow helped them identify missing content and generate goal-aligned revisions (U1, U2, U3, U4). U1 remarked, “The voice assistant tells you what you're missing and what you're not missing,” explaining that this feedback supported a more holistic view of her draft and made it easier to reflect on whether each section served its intended purpose. Similarly, U5 noted that although she might not typically initiate self-evaluation, WriteFlow “actively prompts you to do evaluations,” increasing her awareness of goal completion and progress monitoring.

U4 noted the system helped her “grasp the blueprint of the whole essay” and become “more conscious about whether what I’m writing actually aligns with my goals, and what role it plays in the whole essay.”

Participants who previously described their writing as "divergent" or "unstructured" (U3, U11) expressed that WriteFlow’s goal-oriented workflow aligned more closely with the structured approaches of their co-writers or supervisors, which they aspired to adopt. Across participants, many emphasized that clarifying goals and overall structure helped them maintain focus on their intended direction throughout the writing process (U2-U8, U11, U12).

At the same time, some participants reported cognitive and motivational challenges associated with AI-generated evaluations. U1, U7, and U12 described feeling fatigued by lengthy descriptive feedback, while U12 questioned the AI’s ability to accurately evaluate her writing, noting that her personal standards often exceeded what AI could assess. U7 expressed a preference for constructive and supportive feedback over overly critical evaluations. Rather than receiving long textual critiques, U7 and U12 preferred the AI to provide high-quality worked examples that they could use for comparison and adaptation.
Finally, U3 and U4 cautioned that single-perspective or overly complete responses could constrain divergent thinking and limit deeper exploration. To preserve writer agency and support reflective thinking, they suggested that WriteFlow should offer multiple perspectives or alternative options, encouraging users to critically evaluate suggestions rather than passively accept them.

\subsection{Writer Agency and Critical Engagement with AI Feedback}
\textbf{Preserve Writer Voice and Meaning (R4)}. Participants described WriteFlow as offering a greater sense of authorial control, referring to the extent to which users retain agency over the content and direction of their writing when interacting with AI. Several participants commented that WriteFlow helped them make more informed decisions by providing explanations behind each response (U4, U5). For example, as U5 highlighted: ``This tool allows decision-making at every decision point. When the AI provides something that’s incorrect or off-track, it lets the user make a direct and convenient choice, like accept or reject\ldots It gives you a comparison and a framework for evaluation. After evaluating, then I can decide whether to add that goal to the list.''


Participants (U2, U3, U4) also shared that WriteFlow supported ongoing evaluation of whether their writing stayed aligned with their original intentions, as shared by U2 ``I think I would prefer to use this system to help me to maintain my intention in the writing''. U2 and U3 perceived WriteFlow as a supportive writing partner, one that encouraged them to propose confusion, disagreement, and alternative statements. This seemed to shift them from passively receiving AI-generated content (as with ChatGPT) to actively negotiating meaning and structure, thereby reinforcing their agency and voice throughout writing (U2, U3).

\textbf{Transparent Feedback to Support Critical Evaluation of AI Outputs (R5).} In this study, participants consistently evaluated and verified AI-generated outputs, and this verification process itself stimulated a deeper reflection. This suggests that traceability and transparent reasoning are key not only for supporting informed decision-making, but also for fostering reflective writing practices in human-AI interaction. Six participants reported that the system’s use of concrete, sufficiently detailed evidence and clear, material-grounded reasoning increased the perceived trustworthiness and reliability of its suggestions. U9 further proposed that visually representing the AI’s inference process could better support human verification, noting that tracing this reasoning might also help writers enter a more reflective cognitive state.
U11 emphasized the value of explainable AI, expressing the expectation that the system would “explain why this goal makes sense,” with reference to the source material. 

\section{Discussion}\label{sec12}

This study aimed to design and evaluate WriteFlow, a human-informed, AI-driven writing assistant for reflective academic writing, with the aim of understanding how goal-oriented interaction design can scaffold writers’ metacognitive regulation and foster agency. Using WriteFlow as a design probe in an exploratory Wizard-of-Oz study, we addressed two research questions: (RQ1) how WriteFlow supports metacognitive, goal-oriented writing processes and what design refinements are suggested by user feedback, and (RQ2) how interacting with WriteFlow influences writers’ sense of agency and reliance on AI during academic writing.

Overall, WriteFlow was experienced not merely as a text-generation tool, but as a reflective partner that foregrounded academic writing as an intentional, goal-driven cognitive activity. This aligns with the Cognitive Process Theory of Writing, which conceptualizes writing as a recursive interaction between planning, translating, and reviewing rather than a linear sequence of stages \cite{flower1981cognitive,hayes1986writing}. 

\subsection{Metacognitive Goal-Oriented Writing Process with AI }

Addressing RQ1, the findings show WriteFlow supports metacognitive, goal-oriented writing by making goals explicit, revisitable, and actionable throughout drafting and revision. Consistent with research showing that academic writing is guided by hierarchically structured and evolving goal systems \cite{flower1981cognitive,hayes1986writing}, participants used goals not as fixed plans but as flexible reference points that were continuously refined as new ideas, sources, and feedback emerged, reflecting expert-like writing behavior \cite{hayes1986writing}. Consistent with research showing that specific, proximal, and process-oriented goals enhance metacognition and writing performance \cite{schunk1990goal,reid2013strategy,chung2021impact}, participants reported WriteFlow’s emphasis on articulating and refining goals helped them clarify vague intentions, focus on quality-oriented concerns, and maintain coherence across longer texts. 
Building on prior writing research showing that critically revising rather than passively accepting AI suggestions can foster critical thinking in academic writing \cite{yang2025modifying}, WriteFlow supported reflection on the underlying intentions of writers by linking goals to evolving text and prompting goal achievement's evaluation.

A key contribution relative to prior AI-supported writing systems lies in when metacognitive support is provided. Whereas tools such as VISAR focus on goal construction during planning \cite{zhang2023visar} and systems like Friction and ALure support reflection during revision \cite{zhang2025friction,neshaei2025leveraging}, WriteFlow supports awareness of goal–text alignment during drafting itself, addressing a gap where writers may otherwise fail to notice goal drift in AI-assisted writing.

Features such as goal tracking and goal–text linkage reduced the cognitive load of managing competing objectives, particularly during revision based on reviewer feedback. Participants’ requests for hierarchical goal visualization and adaptive prioritization further highlight the value of making goal structures explicit. This aligns with evidence that planning and quality-oriented goals support higher-level revision \cite{beauvais2011some}, while underscoring the need to balance structure with cognitive load \cite{radwan2024sard}. At the same time, participants noted limits of AI-supported evaluation: lengthy or overly authoritative feedback sometimes caused skepticism. Preferences for worked examples and multiple perspectives suggest that goal-oriented AI support is most effective when it scaffolds reflection and decision-making rather than prescribing solutions.

\subsection{Enhanced Writer Agency and Critical Engagement with AI} 

Addressing RQ2, the findings show that WriteFlow influenced the writer's sense of agency by repositioning AI as a responsible, negotiable partner instead of an authoritative content generator. This is particularly salient given concerns raised in the formative study about overreliance, authorship, and ownership when using general-purpose AI tools (Section 3). Participants described WriteFlow as supporting their ability to preserve writer voice and meaning by requiring explicit decisions, such as accepting, rejecting, or revising goals and suggestions, at key points in the writing process. Participants contrasted this interaction style with experiences of using ChatGPT in a more passive or efficiency-oriented manner. In WriteFlow, agency was reinforced through goal-based framing of AI feedback and through explanations that made the system’s reasoning visible. This transparency enabled participants to critically evaluate AI outputs against their own intentions, supporting informed decision-making rather than deference. This aligns with participants’ expectations for explainable AI and traceable reasoning, as articulated in design requirement 5 (R5).

Importantly, verification and critique of AI outputs were not perceived as friction, but as productive moments of reflection. Participants reported that grounding AI feedback in concrete evidence from their own text increased trust while simultaneously encouraging scrutiny. Suggestions to visualize the AI’s inference process further indicate that transparency may serve not only trust calibration, but also reflective engagement. However, participants also cautioned that single-perspective or overly polished AI outputs could constrain divergent thinking. Their preference for alternative options and multiple viewpoints underscores that preserving agency involves maintaining epistemic openness, not merely control. In this sense, WriteFlow was experienced as a dialogic partner that supported \textit{reflection-in-action }\cite{schon2017reflective}, helping writers navigate uncertainty, evolving goals, and competing constraints in academic writing.

This study has several \textbf{limitations} and \textbf{implications} for the design of AI-supported academic writing tools. \textit{First}, while the Wizard-of-Oz setup enabled fine-grained exploration of interaction dynamics, the use of researcher-prepared materials may have limited participants’ sense of ownership. Future work should examine how goal-oriented AI support functions when writers engage with self-selected topics and authentic writing contexts over longer periods. 
\textit{Second,} participants’ feedback suggests that prompt design is not a purely interface-level concern, but a central mechanism shaping metacognitive engagement. Fixed prompt structures and uniform response lengths constrained reflective depth, indicating a need for adaptive prompting that responds to writers’ expertise, confidence, and stage in the writing process. \textit{Finally}, the involvement of participants with HCI expertise limits generalizability. Future studies should examine how writers from diverse disciplinary, linguistic, and educational backgrounds appropriate goal-oriented AI support, particularly given differences in writing conventions and self-regulatory practices.

In sum, this work demonstrates how LLM-based writing tools can be designed to support metacognition and self-regulated learning by foregrounding iterative goal setting and goal–text alignment throughout the writing process. Rather than optimizing for efficiency alone, WriteFlow illustrates a design direction in which AI systems scaffold reflective dialogue and evolving goal structures, reinforcing writer agency and intentional engagement in the academic writing process.

\begin{credits}
\subsubsection{\ackname} We thank all the study participants for their engagement in this study. The work has in part been supported by the STINT grant: “Capitalizing on  the potentials of technology to promote self-regulation” (MG2018-7984).
\end{credits}


%
%
%
\bibliographystyle{splncs04}
%




\bibliography{sn-bibliography}
\end{document}